\documentclass[11pt]{article}

\usepackage[dvips]{graphicx}

\def\bc{\begin{center}}                \def\ec{\end{center}}
\def\beq{\begin{equation}}              \def\eeq{\end{equation}}
\def\bear{\begin{eqnarray}}            \def\eear{\end{eqnarray}}
\def\bt{\begin{tabular}}               \def\et{\end{tabular}}
\def\dst{\displaystyle}
\def\la{\langle}     \def\ra{\rangle}   
       \def\lb{\label}     
\def\hs{\hspace}      \def\vs{\vspace}
\def\ti{\tiny}        \def\sm{\small}       
  \def\rar{\rightarrow}
       \def\pr{\prime}
   
\def\fns{\footnotesize}

\def\a{\alpha}      \def\bet{\beta}

\def\vphi{\varphi}                  
\begin{document}

\title{ {\hfill{\sm E-print quant-ph/0410045}}\\[5mm]
\bf On the `Polarized distances between quantum states and  observables'}
\author{D.A. Trifonov \\
        Institute for Nuclear Research, Sofia, Bulgaria}
\date{\today}
\maketitle

\begin{abstract}
The scheme for construction of distances, presented in our previous paper
quant- ph/0005087, v.1 (Ref. 1) is amended. The formulation of Proposition
1 of Ref. 1 does not ensure the triangle inequality, therefore some of the
functionals $D(a,b)$ in Ref. 1 are in fact quasi-distances.  In this note
we formulate sufficient conditions for a functional $D(a,b)$ of the
(squared) form $D(a,b)^2 = f(a)^2 + f(b)^2 - 2f(a)f(b)g(a,b)$ to be
a distance and provide some examples of such distances. A one parameter
generalization of a bounded distance of the (squared) form
$D(a,b)^2 = D_0^2 (1 - g(a,b))$, which includes the known Bures-Uhlmann
and Hilbert-Schmidt distances between quantum states, is established.
\end{abstract}

In the scheme of paper quant-ph/0005087 [1] (to be cited also as Ref. 1)
two functionals,  $f(a)$ and $g(a,b)$ on a set ${\cal A}$ ($a,b
\in {\cal A}$) are involved.  $f(a)$ was required to be positive,
$g(a,b)$  -- symmetric, $g(a,b) = g(b,a)$, with values in the interval
$[-1,1]$. In addition it was supposed that $a=b$ is equivalent to $g(a,b)=1$
and $f(a)=f(b)$ (i.e. $g(a,b)=1,\,  f(a)=f(b)\,\, \longleftrightarrow\,\,
a=b$). Then the expression $D[a,b]$,
\beq\lb{D}
D[a,b] = \left(f(a)^2 + f(b)^2 - 2f(a)f(b)\,g(a,b)\right)^{1/2}
\eeq
was proposed [1] as distance between elements of ${\cal A}$. The
functional $g(a,b)$ is a {\it cosine-type} functional, and $f(a)$ plays
the role of {\it polarization}. If $f(a)=$ const. then the distance (\ref{D})
is {\it not polarized}.  In somewhat different form the notion of polarized
distance was introduced in [2].
\vs{2mm}

{\sm
Before proceed further let us recall the defining properties (d1) - (d4) of
the {\it distance} $D(a,b)$ between elements $a,\,b,\,c$ of a given set
${\cal A}$:\\[-3mm]
$$ D(a,b) \geq 0\quad {\mbox{\fns (nonnegativity)}}\,,
\\[-1mm] \eqno{\rm(d1)}$$
$$ D(a,b) = 0\quad {\rm iff}\,\, a = b\quad
{\mbox{\fns (Euclidean\,\, property)}},\\[-1mm] \eqno{\rm(d2)} $$
$$D(a,b) = D(b,a)\quad {\mbox{\fns (symmetricity)}}\,,\\[-1mm]
\eqno{\rm(d3)} $$
 $$ D(a,b) + D(b,c) \geq D(a,c)\quad {\mbox{\fns (triangle\,\, inequality)}}.
\eqno{\rm(d4)}$$

Trivial distance $D(a,a)=0$, $D(a,b\!\neq\! a)=1$ always exists.
If all but (d2) are valid then $D(a,b)$ is called {\it pseudo-distance}.
If all but the triangle inequality (d4) are valid then $D(a,b)$ would be
called {\it quasi-distance}.}
\vs{2mm}

However the above requirements on $g(a,b)$ in Eq. (\ref{D}) do not ensure
the triangle inequality (d4). Some of the "new distances" $D[a,b]$ of the
type (\ref{D}), constructed in Ref. 1 (the expressions $D_{\rm III}$ in
subsection 5.1.1 of [1] for example) violate (d4), and are in fact
{\it quasi-distances} only. The expressions $D^\pr[A,B]$ of [1] should be
regarded as quasi-distances between quantum observables $A$ and $B$.

In this note we accomplish the Proposition 1 of [1] adding an extra
requirement on the functional $g(a,b)$ in (\ref{D}) in order to ensure the
triangle inequality. The Bures-Uhlmann (BU) distance between pure quantum
states is an example that obeys these additional requirements.  Then we
provide a one parameter generalization of an angle distance
$\arccos(g(a,b))$ (Proposition 2) that entails a one parameter
generalization of bounded distances of BU type between pure quantum
states. The generalization includes the Hilbert-Schmidt (HS) distance
between pure states and also obeys the conditions of Proposition 1.
Finally we note that some of the "new polarized distances" in [1] (those
based on the HS and BU distances) do satisfy (d4) and are correct
distances on the corresponding set of quantum states.

The  distance proposition (a sufficient condition for a
distance of the type (\ref{D})) should have the following form:
\vspace{2mm}

{\bf Proposition 1.} {\sl Let $f(a)$ be a positive functional on a set
${\cal A}$, and $g(a,b)$ be a symmetric functional on ${\cal A}\times{\cal
A}$ ($a,b \in {\cal A}$) with the property\vs{-3mm}
\beq\lb{prop1}
|g(a,b)| \leq 1. \vs{-3mm}
\eeq
Then the expression \vs{-1mm}
\beq\lb{Da}
D(a,b;f,g) = \left(f(a)^2 + f(b)^2 - 2f(a)f(b)\,g(a,b)\right)^{1/2}
\eeq
is a distance between elements of ${\cal A}$ if
$g(a,b) = g(b,a)$ and
\vs{-1mm}\beq\lb{prop2}
f(a) = f(b),\, g(a,b) = 1\quad \longleftrightarrow\quad a=b,
\eeq
\vs{-4mm} \beq\lb{prop3}
|\varphi_{ab} - \varphi_{bc}|\, \leq\, \varphi_{ac}\,\leq\, {\rm
min}\{\varphi_{ab} + \varphi_{bc},\,\, 2\pi - (\varphi_{ab} + \varphi_{bc})\},
\eeq
where $\varphi_{ij} = \arccos(g(i,j)),\quad i,j=a,b,c$}.
\vs{2mm}

{\sl Proof}. The first three required properties for $D(a,b;f,g)$ as a
distance are easily seen to follow straightforwardly from eqs.
(\ref{prop1}) and (\ref{prop2}), and the inequality $(f(a)-f(b))^2 \geq 0$.
To prove the triangle inequality we consider a
reper in ${\mathbf R}_3$, consisting of three unit vectors $\vec{e}_a $,
$\vec{e}_b$ and $\vec{e}_c$  (see figure 1), the angles subtended by the
three pairs of unit vectors
\bc
\hs{5mm}\bt{ll} \makebox(60,50)[tt]{\input{figure1.pic}}&\hs{20mm}
\makebox(50,40)[rc]{ \begin{minipage}{47mm}
\vs{-5mm} {\small
\vs{7mm}{\bf Figure 1.} A distance triangle $ABC$ (thick sides) of the form
(\ref{Da}): $AB  = D(a,b;f,g)$, $BC  = D(b,c;f,g)$,\,
$CA  = D(c,a;f,g)$. The Points $O,\,A$, $B,\,C^\pr$ are in the plane $YOZ$.
The distance triangle is a base of a pyramid with apex in the origin $O$ and
edges $OA=f(a),\, OB=f(b),\, OC=f(c)$. }
\end{minipage}}
\et
\ec
$(\vec{e}_i,\vec{e}_j)$, $i,j = a,b,c$, being equal to
$\arccos(g(i,j))\equiv \vphi_{ij}$ respectively.

We first put the vectors $f(a)\vec{e}_a=\vec{OA}$ and $f(b)\vec{e}_b =
\vec{OB}$, with the angle between them equal to  $\arccos(g(a,b))\equiv
\vphi_{ab}$. Next we put (in the plane $YOZ$ of $\vec{OA}$ and $\vec{OB}$)
the vector $f(c)\vec{e}_{c^\pr}=\vec{OC^\pr}$ with the
angle between $OB$ and $OC^\pr$ equal to $\vphi_{bc}$ and, if
$\vphi_{ac}$ is less than $\vphi_{ab}+\vphi_{bc}$ (as it is the case shown
on figure 1, where $\vphi_{ab}+\vphi_{bc} > \pi$), rotate $\vec{OC^\pr}$
around $\vec{OB}$ until the angle between $OC^\pr$ and the already
fixed $OA$ becomes equal to $\vphi_{ac}$.
The required property (\ref{prop3}) ensures this possibility.  The final
position of $\vec{OC^\pr}$ is $\vec{OC}$. In this way we obtain the
triangle $ABC$ (thick lines), the side of which are exactly the three
distances $D(a,b;f,g)$, $D(b,c;f,g)$ and $D(b,c;f,g)$ of the form
(\ref{Da}). And the sides of a triangle do satisfy the triangle
inequality. End of Proof.
\vs{2mm}

{\sl Remark}. If the cosine functional obey the inequalities $0\leq
g(a,b)\leq 1$, then  all the three distance angles $\vphi_{ij}$ are in the
interval $[0,\pi/2]$, and the condition (\ref{prop3}) simplifies to
\beq\lb{prop3a}
|\varphi_{ab} - \varphi_{bc}| \leq \varphi_{ac}\leq
\varphi_{ab} + \varphi_{bc}.
\eeq
The distance triangle pyramids in these cases are in the first cartesian
octant.
\vs{2mm}

Important known examples of distances of the form (\ref{D})
are the norm distance between vectors in Hilbert spaces and the
Hilbert-Schmidt (HS) and the Bures-Uhlmann (BU) distances $D_{\mbox{\tiny
HS}}$, $D_{\mbox{\tiny BU}}$ between quantum states.  The HS and BU
distances read

 \beq\lb{HS}
\bt{r}
$\dst D_{\mbox{\tiny HS}}(\rho_1,\rho_2) =
\sqrt{f_{\mbox{\tiny HS}}^2(\rho_1) + f_{\mbox{\tiny HS}}^2(\rho_2) - 2
f_{\mbox{\tiny HS}}(\rho_1)f_{\mbox{\tiny HS}}(\rho_1)
g_{\mbox{\tiny HS}}(\rho_1,\rho_2)}$,\\[3mm]
$\dst f_{\mbox{\tiny HS}}(\rho) = \sqrt{{\rm Tr}(\rho^2)},\quad
g_{\mbox{\tiny HS}} = {\rm Tr}(\rho_1\rho_2)/
\sqrt{{\rm Tr}(\rho_1^2){\rm Tr}(\rho_2^2)},$
\et
\eeq
\vs{0.5mm}

\beq\lb{BU}
D_{\mbox{\ti BU}}(\rho_1,\rho_2) =
\sqrt{2(1- g_{\mbox{\ti BU}}(\rho_1,\rho_2))}, \quad
g_{\mbox{\ti BU}} = {\rm Tr}\sqrt{\sqrt{\rho_1}\rho_2\sqrt{\rho_1}},
 \eeq
where $\rho$ is the density operator, representing quantum states. One
sees that
$$ 0 \leq g_{\mbox{\ti BU}} \leq 1,\quad  0\leq g_{\mbox{\ti HS}}\leq 1. $$

For pure states $\rho=|\psi\ra\la\psi|$ these formulas simplify to
 \beq\lb{HSa}
D_{\mbox{\tiny HS}}(\psi_1,\psi_2) =
\sqrt{2(1- g_{\mbox{\tiny HS}}(\psi_1,\psi_2))}, \quad
g_{\mbox{\tiny HS}}(\psi_1,\psi_2) = |\la \psi_1|\psi_2\ra|^2,
\eeq
\beq\lb{BUa}
D_{\mbox{\ti BU}}(\psi_1,\psi_2) =
\sqrt{2(1- g_{\mbox{\ti BU}}(\psi_1,\psi_2))}, \quad
g_{\mbox{\ti BU}}(\psi_1,\psi_2) = |\la \psi_1|\psi_2\ra|,
 \eeq
where $\la\psi_1|\psi_2\ra$ is the scalar product.

We shall show below that for pure states
$g_{\mbox{\ti BU}}$ and $g_{\mbox{\ti HS}}$ obey (\ref{prop3a}), i.e.
the HS and BU distances between pure states are examples of known distances
which obey the requirements of Proposition 1. (It is not clear wether
$g_{\mbox{\ti BU}}$ and $g_{\mbox{\ti HS}}$ obey the angle condition
(\ref{prop3a}) in the case of mixed states).

The proof for $g_{\mbox{\ti BU}}(\psi_1,\psi_2)$  is very simple.  It
resorts to the known fact that the angle $\vphi_{12} = \arccos(|\la\psi_1 
|\psi_2\ra|)$ (the Hilbert space angle) is a distance between pure quantum
states [3]. Therefore  $\vphi_{12}+\vphi_{23} \geq \vphi_{31}$, what
exactly is the requirement (\ref{prop3a}) for $g_{\mbox{\ti BU}}
(\psi_1,\psi_2)$. The property (\ref{prop3a}) for $g_{\mbox{\ti HS}}
(\psi_1,\psi_2)$ follows from the Proposition 2 below.

Let us note that not every distance of the form (\ref{D}) obeys the
property (\ref{prop3}), in particular not every finite distance
of the form
\beq\lb{D_0}
D(a,b) = D_0\sqrt{1-g(a,b)},\quad 0\leq g(a,b)\leq 1
\eeq
obeys (\ref{prop3a}) (One can easily find counter-examples numerically).
In fact any finite distance $D(a,b)\leq D_0$ can be put in
the form (\ref{D_0}) with $g(a,b) = 1 - D^2(a,b)/D_0^2$.
\vs{3mm}

{\bf Proposition 2}. {\sl If the angle $\vphi(a,b) = \arccos(g(a,b))$ is a
distance, $0\leq \vphi(a,b) \leq \pi/2$, on a set ${\cal A}$, then}
\beq\lb{vphitau}
\arccos(g^\tau(a,b)) \equiv \vphi(a,b;\tau)
\eeq
{\sl is also a distance for $\tau > 1$.}
\vs{2mm}

{\sl Proof.} It is clear that we have to prove the triangle
inequality for $\vphi(a,b;\tau)$ only, since the properties (d1)-(d3) of
$\vphi(a,b;\tau)$ straightforwardly follow from those of $\vphi(a,b) =
\vphi(a,b;1)$.

In the notations $x=g(a,b)$, $y=g(b,c)$, $z=g(a,c)$ (note that $0\leq
x,y,z \leq 1$) the triangle inequality for $\vphi(a,b)$ reads
\beq\lb{3ineq 1}
\vphi(x)+\vphi(y) \geq \vphi(z)\,.
\eeq
We have to prove that
\beq\lb{3ineq t}
\vphi(x,\tau)+\vphi(y,\tau) \geq \vphi(z,\tau)\,
\eeq
for $\tau > 1$.

The proof of (\ref{3ineq t}) resorts to the monotone character (with respect
to the parameter $\tau$) of function $\vphi(x,\tau)$  and its first and
second derivatives $\vphi^\pr(x,\tau)=d\vphi(x,\tau)/d\tau$ and
$\vphi^{\pr\pr}(x,\tau)=d^2\vphi(x,\tau)/d\tau^2$: $\vphi(x,\tau)$
increases from $0$ at $\tau=0$ to $\pi/2$ at $\tau\rar\infty$
(see three dashed lines in figure 2, where $x=0.9,\,0.6,\,0.2$);
$\vphi^\pr (x,\tau)$ is strictly decreasing function of $\tau$, $\infty
> \vphi^\pr (x,\tau) > 0$,  and $\vphi^{\pr\pr} (x,\tau)$ is strictly
increasing negative function of $\tau$, $-\infty < \vphi^{\pr\pr} (x,\tau)
< -0$. The sum $\vphi(x,\tau)+\vphi(y,\tau)\equiv \vphi_+(x,y;\tau)$ is also
strictly increasing function of $\tau$ (with strictly decreasing positive first
derivative) and tending to $\pi$ at $\tau \rar \infty$: At $\tau >> 1$ one
has $\vphi_+(x,y;\tau) > \vphi(z;\tau)$.
\vs{2mm}

\bc
\hs{5mm}\bt{ll} \includegraphics[width=65mm,height=45mm]{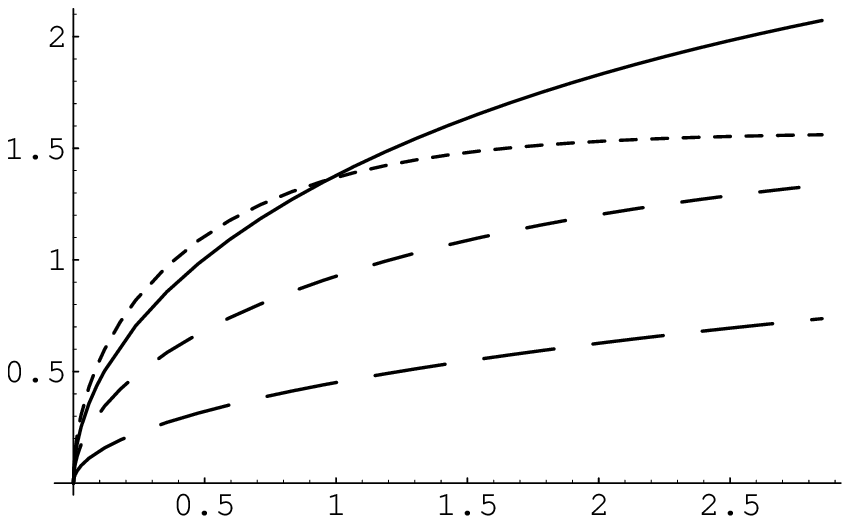}$\qquad$&
\makebox(50,40)[rc]{\begin{minipage}{45mm}
{\small \vs{7mm}{\bf Figure 2.}
$\arccos(x^t)$ as a function of $t$ for $x=0.9$ (the lower dashed line),
$x=0.6$ (the intermediate dashed line) and $x=0.2$ (the top dashed line).
The solid line represents the sum $\arccos(0.9^t)+\arccos(0.6^t)$.}
\end{minipage}}
\et
\ec\vs{3mm}
Due to these
monotone properties the two curves, $\vphi_+(x,y;\tau)$ and $\vphi(z;\tau)$,
either do not intersect (i.e. $\vphi_+(x,y;\tau) > \vphi(z;\tau)$ for all
$\tau>0$ ) or intersect at one point only, after which $\vphi_+(x,y;\tau)
> \vphi(z;\tau)$. In view of (\ref{3ineq 1}) the intersection can
occur at $t\leq 1$ only (see figure 2, where the solid line represents
$\vphi(0.9,0.6;\tau)$). Two or more intersections mean that
$\vphi_+(x,y;\tau)$ is oscillating around $\vphi(z;\tau)$ which is
impossible due to strictly monotone behavior of these functions and their
derivatives. End of the Proof.

\vs{2mm}

Next we note that if $\vphi(a,b;\tau)$ is a distance it does satisfy the
inequalities (\ref{prop3a}). As a result we have the \vs{2mm}

{\bf Corolarry 1.} {\sl If $D(a,b)$ is a bounded distance of the form
(\ref{D_0}) satisfying (\ref{prop3a}) then $D(a,b;\tau)= D_0\sqrt{1-g^\tau
(a,b)}$ is also a distance satisfying (\ref{prop3a}). }\vs{2mm}

The BU distance (\ref{BUa}) between pure quantum states is of the form
(\ref{D_0}) with angle (the Hilbert space angle [3]) $\vphi(\psi_1,\psi_2)
= \arccos(|\la \psi_1|\psi_2\ra|)$ satisfying (\ref{prop3a})
($\vphi(\psi_1,\psi_2) = \arccos(g_{\mbox{\ti BU}}(\psi_1,\psi_2))$) .
Therefore the functionals $D(\psi_1,\psi_2;\tau)$,
\beq\lb{BU t}
D(\psi_1,\psi_2;\tau) = \sqrt{2(1-|\la\psi_1|\psi_2\ra|^\tau)},\quad
\tau \geq 1,
\eeq
are  also  distances between pure quantum states, satisfying all the
requirements of the Proposition 1 (with $f(a) = 1$ and the stronger
inequalities (\ref{prop3a})). Any distance triangle with sides
$D(\psi_1,\psi_2;\tau)$, $D(\psi_2,\psi_3;\tau)$ and
$D(\psi_3,\psi_1;\tau)$ is a base of (slant) pyramid with unit edges. It
can be realized as a triangle between three points on the first octant of
the unit sphere, the apex of the pyramid being placed in the center of the
sphere.
At $\tau =2$ the functional $D(\psi_1,\psi_2;\tau)$ coincides with the HS
distance between pure states.

It is worth noting another simple generalization: if $D(a,b)$ is a bounded
distance on a set ${\cal A}$ of the form (\ref{D_0}) (the condition
(\ref{prop3a}) is not required) then
\beq\lb{D t}
D(a,b;\tau) = D_0\sqrt{1-g^\tau(a,b)}
\eeq
is also a distance for $\tau \geq 1$. The proof of the statement resorts on
the strictly monotone character of $\sqrt{1-x^\tau}$  and its first and
second derivatives as functions of $\tau$ and is carried out similarly to
the case of Proposition 2. It may be useful also to note that at $\tau
\rar \infty$ the expression $D(a,b;\tau)$ tends to the trivial distance in
${\cal A}$: $D(a,b;\tau\rar\infty) = 1$, if $a\neq b$ and $D(a,b;
\tau\rar\infty) = 0$ if $a=b$.
\vs{2mm}

The functionals $D_{\rm Ia}$ and $D_{\rm Ib}$ of subsection 5.1.1 of
[1] are of the form (\ref{D}) with $g = g_{\mbox{\tiny HS}}$. Therefore,
in view of the Proposition 2 they obey the conditions
(\ref{prop1}), (\ref{prop2}) and (\ref{prop3a}) and are polarized distances
between pure quantum states (note however that in Eqs. (57), (58) of [1]
the numerator $1\!-\!\exp(-|\a\!-\!\bet|^2)$ should be replaced by
$\exp(-|\a\!-\!\bet|^2)$). The expressions $D^\pr[A,B]$ of [1] may satisfy
the triangle inequality on some restricted subsets only, and thus should be
regarded as quasi-distances between quantum observables $A$ and $B$. The
expressions $D_{III}(\rho_1,\rho_2;B,\mu,\nu)$ [1] would be a (polarized)
quasi-distance between (mixed) states if Tr$(\rho_1\rho_2)$ is replaced by
Tr$(\sqrt{\rho_1} \rho_2\sqrt{\rho_1})^{1/2}$.

Let us note at this point, that quasi-distances are also involved in some
physical applications. Thus in the measure of entanglement
$E({\sigma}):= \min_{ \rho \in \cal D}\,\,\, D( \sigma ||\rho)$ [4] the
functional  $D$ is supposed to be "any measure of {\em distance} between
the two density matrices $\rho$ and $ \sigma$ such that $E({\sigma})$
satisfies the above three conditions" [4]. In particular it
could be of the form $D_B({\sigma}||{\rho}) = 2- 2\sqrt{F(\sigma,\rho)}$,
where $F(\sigma,\rho):= \left[{\rm Tr}\{\sqrt{{\rho}}
{\sigma}\sqrt{{\rho}}\}^{1/2} \right]^2$ [4].  The functional
$D_B({\sigma}||{\rho})$ coincides with the squared BU distance (\ref{BU}),
and squared distances are quasi-distances -- they obey (d1)-(d3) but
violate the triangle inequality. (The other choice $S( \sigma ||\rho)$
used in [4] for $D( \sigma ||\rho)$ is not even a quasi-distance, since it
is not symmetric).  
\vs{2mm}

{\large \bf Conclusion.} We have established sufficient conditions for
distances of the form (\ref{Da}) on a given set ${\cal A}$ (Proposition 1)
and showed that the known Bures-Uhlmann (BU) and Hilbert-Schmidt (HS)
distances between pure quantum states, and some of the new distances in
Ref. 1, obey the required conditions (\ref{prop3}), or (\ref{prop3a}).
This is an amended version of the incomplete Proposition 1 of [1].  We
have also constructed a one parameter generalization
$D(\psi_1,\psi_2;\tau)$, Eq. (\ref{BU t}), of BU and HS distances between
pure quantum states that also obeys the requirements of Proposition 1.
Intermediately we showed  that $\arccos(|\la\psi_1|\psi_2\ra |^\tau)$ is a
distance between pure quantum states for $\tau \geq 1$. The latter is a
particular case of Proposition 2.  The same one parameter generalization
(eq. (\ref{D t})) is extended to bounded distances of the form (\ref{D_0})
without requirement of property (\ref{prop3a}) for the symmetric 
functional $g(a,b)$.

The established generalizations can be combined (according to
(\ref{Da})) with any positive functional $f(a)$ on ${\cal A}$ to obtain
new (polarized, therefore more sensitive) distances between elements of
${\cal A}$, in particular between states of quantum systems.  Let us
remind that distances  can be used as measures of distinguishability
between elements of ${\cal A}$. In particular the notion of
distinguishability between quantum states has shown to play an important
role in the modern quantum information theory.

\end{document}